\documentclass[review]{elsarticle}

\journal{Journal of \LaTeX\ Templates}

\begin{document}

\begin{frontmatter}

\title{A study of Variability of the Marginal Am star HD\,176843 observed in the  {\it Kepler} field}


\author[l1,l2]{C. Ulusoy}
\author[l3]{I. Stateva}
\author[l4]{B. Ula{\c s}}
\author[l5]{F. Ali{\c c}avu{\c s}}
\author[l3]{I.Kh.Iliev}
\author[l3]{M. Napetova}
\author[l2,l6]{E.Kaygan}

\address[l1]{School of Aviation, Girne American University, University Drive, PO Box 5, 99428 Karmi Campus, Karaoglanoglu,Girne (Kyrenia), Cyprus}
\address[l2]{College of Graduate Studies, University of South Africa, PO Box 392, Unisa, 0003, Pretoria, South Africa}
\address[l3]{Institute of Astronomy with NAO, Bulgarian Academy of Sciences,blvd.Tsarigradsko chaussee 72, Sofia 1784, Bulgaria}
\address[l4]{{I}zmir Turk College Planetarium, 8019/21 sok., No: 22, \.{I}zmir, Turkey}
\address[l5] {Department of Physics, Faculty of Arts and Sciences, {\c C}anakkale Onsekiz Mart University, 17100 {\c C}anakkale, Turkey}
\address[l6] {College of Engineering, Design and Physical Science, Brunel University, Uxbridge, UB8 3PH, UK}




\begin{abstract}
We present results of a study of variability of the marginal Am star HD\,176843  observed in the {\it Kepler} field. {\it Kepler} photometry and ground--based spectroscopy are used to investigate the light variations of the star. HD\,176843 is classified as a marginal Am star that shows $\delta$\ Sct type pulsations. From an analysis of the {\it Kepler} time series, we find that the light curve of HD\,176843 is dominated by three modes with frequencies $f_{1}$=0.1145, $f_{2}$=0.0162 and  $f_{3}$=0.1078 d$^{-1}$. The amplitude of the radial velocity variations of about 10 km/s is much more than the radial velocity errors and allows us to conclude clear radial velocity variations. Using the radial velocity data and the adopted spectra, the orbital solution of HD\,176843 is also obtained with an orbital period of 34.14 days. However, the available photometric data show no significant evidence for any possible motion in the binary system.
\end{abstract}

\begin{keyword}
\texttt{stars; variables; stars; oscillations (including pulsations) stars; individual}
\end{keyword}

\end{frontmatter}


\section{Introduction}
The classical Am (metallic--lined) stars are those whose spectrum shows relatively strong absorption of metallic lines (Iron or Iron-group elements) and relatively weak Calcium (Ca) and/or Scandium (Sc) lines that appear between A- and early F spectral types \citep{tmo40,rom48}. They are organized in subclasses corresponding to the K line, Hydrogen and metallic lines respectively. Thus, Am stars are generally classified into two groups: classical Am stars (Am) and marginal Am stars (Am:).
The spectral types of the classical Am stars are determined from the CaII K line, Hydrogen and metal lines corresponding to five and more spectral subclasses (e.g., kA0hA0VmA1). On the other hand, the marginal Am stars show fewer spectral subclasses between the CaII K and metallic lines with milder abundance anomalies unlike the classical ones.

The Am stars are known to be slow rotators \citep{sbak54, sbak55} which causes abundance anomalies (if $v\sin\,i< 120$ d $^{-1}$) due to interaction between gravitational settling and radiative levitation in the presence of absent or weak magnetic field.
In this case, meridional turbulent motions due to slow rotation allow for He to settle from the partial He II ionization zone gravitationally, and therefore $\delta$\ Sct type pulsations are not expected to be driven by the $\kappa$\ mechanism in these stars \citep{bre70, bag73, kurtz76,aerts2010}.

The Am stars were historically known as a class of non-pulsating variables \citep{bre70,kurtz76}. However, pioneering space missions such as {\it MOST} \citep{wal03}, {\it CoRoT} \citep{bag06} and, {\it Kepler} \citep{bor10} provided that many hybrid type pulsator, which shows more than one frequency regime (e.g., pressure and gravity modes), candidates with very low amplitudes can be detected with high-precision techniques. Moreover, using the data of A stars from {\it Kepler} \cite{gri10} showed that both high and low frequencies could be observed in all stars that lie at the $\delta$\ Sct instability strip. The recent study presented that around 200 of known 1600 Am stars are found to be pulsating $\delta$\ Sct and $\gamma$\ Dor stars \citep{sly11, sly17}. 
Some marginal Am stars are also found at the red edge of the instability strip  \citep{sly11, sly17, caba12} rather than where the majority of Am stars are located at the blue age, which means that it can be difficult to distinguish them from $\delta$\ Sct stars. These new results have made Am stars unique targets to test the latest diffusion scenario \citep{tur00} regarding their location in the HR Diagram.

The recent results using {\it WASP} photometry show that there are only 11 Am stars found among 249 binaries.\cite{sly14} also reported  that only 4 out of 70 Am stars in eclipsing binary systems have pulsating components detected by {\it WASP}. Furthermore, {\it Kepler} and {\it K2} telescopes  observed 144 Am stars but only 42 of them showed principal pulsation frequencies and amplitudes within the $\delta$\ Sct range. 
In addition, the vast majority of A and Am stars observed with The Large Sky Area Multi-Object Fiber Spectroscopic
Telescope ({\it LAMOST}) and  {\it WASP}  provided us with the opportunity to claim the presence of $\delta$\ Sct pulsations in Am stars which are located to a region close to the cooler edge of the $\delta$\ Sct instability strip within range $6900 K < T_{\rm eff} < 7600 K$ \citep {zhao12,sly17}.

\cite{hou15} also showed that the $\Delta$ index can be used as metallicism (degree of chemical peculiarity) index to separate the marginal Am stars from the classical ones. Accordingly, this value is defined as indicator of the numerical difference in the k and m spectral types. 
\cite{sly17} presented that there is a negative correlation between the incidence of pulsations in
Am stars and metallicism. What is more, they found that the maximum amplitude of the pulsations is not significantly related to metallicism. 

A large number of the Am stars have been found to be members of short period binary systems  \citep{abt67,cap07,sly14}. 
This would be a possible explanation for low rotation rates because of rotational braking induced by tidal friction in a close binary system \citep{abmoy73,wol83,sly17}. For those Am stars found in a binary system, [Ca/Fe] indicator can be associated with some of the binary parameters. This indicator was determined as an expression [Ca/Fe] = [Ca/H] - [Fe/H], where [N/H] = $\log(N/H)_* - \log(N/H)_\odot$. Am peculiarities were found to increase with increasing orbital eccentricity but no evidence of such a correlation with orbital period was found \citep{stateva12}.  It should  also be noted that around 60--70 percent of Am stars are suggested to be spectroscopic binaries by\citep{sly14}. Yet, the role of binarity is not clear for the Am phenomenon.

The star, HD\,176843 ( {\it Kepler} ID:KIC \,9204718 , A3mF0, V=7$^{m}$ .51), was first classified as an Am star by  \cite{flo75} and it was recently listed to be a marginal Am  star showing $\delta$\ Sct pulsations within a possible contact binary system \citep{uyt11}. Our study is structured as follows: we first present the results of a detailed examination of {\it Kepler} photometry, including method of data reduction and frequency analyses. Section 4  deals with the spectroscopic observations. Finally, a brief summary and conclusions are reported in the last section.

\section{The {\it Kepler} Photometry}

{\it Kepler} photometry is used to investigate the light variations of HD\,176843. The {\it Kepler} \footnote{\it http://kepler.nasa.gov/}space telescope, designed to detect Earth-like planets around the Sun-like stars within the habitable zone \citep{koh10, bor10}, was launched on 2009 March 6. {\it Kepler} continuously observed a 105 square degree area of the sky targeted in the constellations of Cygnus and Lyra.
HD\,176843 was observed with 1- min exposures in Short--Cadence (SC) mode (only Q3.3) and in Long--Cadence (LC) (29.45 min) modes (Q0-Q17). Since vast majority of the observations were obtained in LC we therefore analyzed the LC data collected between the {\it Kepler} commissioning quarters Q0 and Q17.
The data are also publicly available on the Barbara A. Mikulski Archive for Space Telescopes \footnote{\it MAST, http://archive.stsci.edu} 

\section{Light Curve Analyses}
In order to perform the analyses, cotrending basis vector (CBV) files were first applied to data for cotrending  the simple aperture Photometry (SAP) fluxes and removal of instrumental systematics from the light curve by using {\tt kepcotrend} task of {\tt PyKE} package \citep{sti12}. We converted the raw SAP fluxes to magnitudes by using the formula $-2.5\log F$. The magnitudes were also corrected to zero mean by removing a linear trend. To remove the linear trend in the light curves, we first represented the each quarter dataset by a linear equation, individually. Then we subtracted the equation from each data point of the related quarter.
\subsection{Frequency Analysis}
\begin{figure*}
	\centering
	\includegraphics{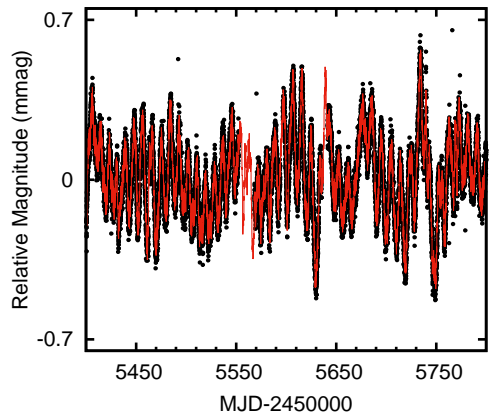}
	\caption{The data between 2455400 and 2455800 are plotted with the analysis result.}
	\label{fitfig}
\end{figure*}

\begin{figure*}
	\centering
	\includegraphics[scale=0.85]{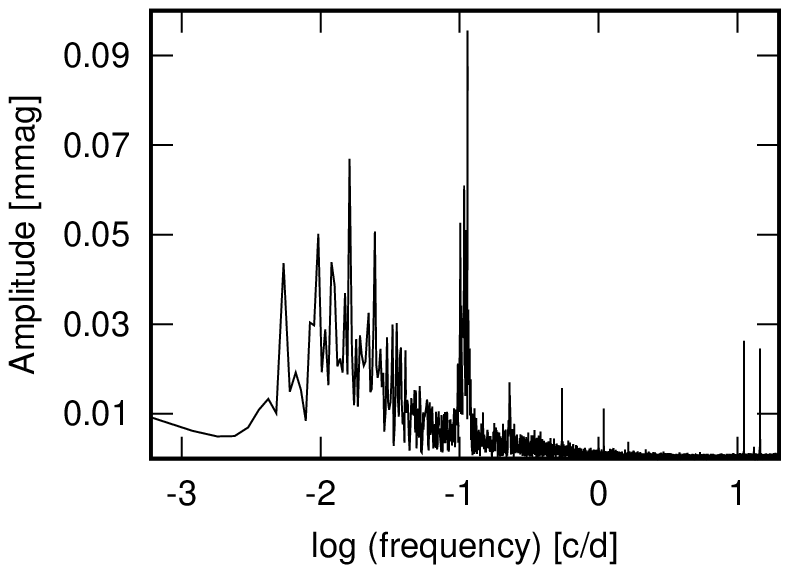}
	\caption{Frequency spectrum yielded by the analysis.}
	\label{periodogr}
\end{figure*}

The frequency extraction of HD\,176843 was obtained by using the {\tt SigSpeC} code \citep{ree07}. The program computes the spectral significance levels for the discrete Fourier transform (DFT) amplitude spectra of time series with arbitrary time sampling.
The default significance threshold in {\tt SigSpeC} is set at 4.1784 that theoretically corresponds to  $S/N$ = 3.5 \citep{ree07, bre11}. The theoretical Rayleigh resolution is $1/T = 0.0007$~d$^{-1}$. Using LC time series, the software package yielded 5 genuine frequencies, 327 combination terms and harmonics up to 5$f_3$.  We also confirm that the light variation is dominated by three frequencies  $f_{1}$=0.1145, $f_{2}$=0.0162 and  $f_{3}$=0.1078 d$^{-1}$ as previously reported by \cite{bal15}. The lowest frequency that appears to be significant is 0.0162 d$^{-1}$. On the other hand, the term with the lowest amplitude is detected to be significant in the region around 14.5363 d$^{-1}$. The resulting frequencies of HD\,176843 are listed in Table \ref{tabfrq}, together with their amplitudes, phases and $S/N$ values and uncertainties. The agreement between analysis and the observation is shown for a certain time interval in Fig.~\ref{fitfig}. The frequency spectrum is also presented in Fig.~\ref{periodogr}.

\begin{table}
\caption{Result of the frequency analysis. Uncertainties are presented in parentheses. We give first ten frequencies having the largest amplitudes for the sake of saving space. The full list of frequencies is available on request from CU.}
\label{tabfrq}
\begin{tabular}{lcccc}
\hline
 &Frequency (d$^{-1}$)& Amplitude (mmag) & Phase & S/N \\
\hline
$f_{1}$ &0.11454(2)&	0.098(2)&	1.29(12)&		68\\
$f_{2}$ &0.01619(2)&	0.073(2)&	4.61(15)&		52\\
$f_{3}$ &0.10782(2)&	0.075(3)&	1.75(16)&		49\\
$f_{4}$ &11.1729(4)&	0.031(2)&	2.19(26)&		31\\
$f_{5}$ &14.5363(4)&	0.026(1)&	6.00(26)&	    31\\
$f_{6}=2f_{1}$&0.22894(5)&0.019(1)&4.98(3)&25\\
$f_{7}=3f_{1}$&0.34426(28)&0.002(1)&3.40(19)&4\\
$f_{8}=2f_{2}$&0.03277(4)&0.034(2)&5.61(3)&29\\
$f_{9}=4f_{2}$&0.06517(11)&0.015(3)&4.18(8)&10\\
$f_{10}=5f_{2}$&0.08111(8)&0.011(1)&2.61(5)&15\\
$f_{11}=7f_{2}$&0.11279(5)&0.083(6)&4.11(3)&25\\
$f_{12}=8f_{2}$&0.12884(8)&0.038(4)&4.83(6)&14\\
$f_{13}=9f_{2}$&0.14546(16)&0.005(1)&1.26(11)&7\\
$f_{14}=2f_{3}$&0.21535(15)&0.007(1)&3.82(10)&8\\
$f_{15}=5f_{3}$&0.53928(25)&0.002(1)&0.64(17)&5\\
\hline
\end{tabular}
\end{table}

\begin{figure*}
	\centering
	\includegraphics{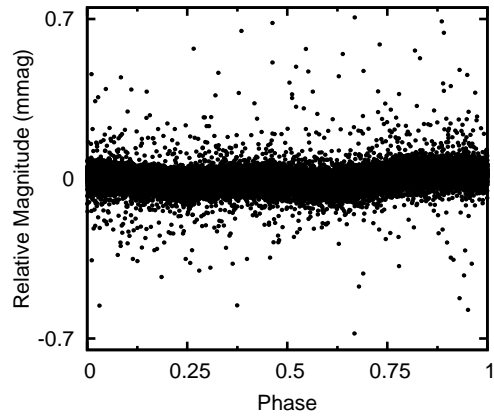}
	\caption{All pre-whitened data except 0.5474 d$^{-1}$ and its harmonics are folded in 1.8268 days.}
	\label{resifig}
\end{figure*}

We also analyzed the data covering all LC quarters (Q0-Q17) to test the binary influence on the light variation suggested by \cite{bal15}. Thus, we pre-whitened all frequencies from the data except 0.5474 d$^{-1}$ and its harmonics (1.094  d$^{-1}$ and 1.642  d$^{-1}$) which is suggested as the binary period. However, the resulting curve folded in 0.5474 d$^{-1}$ (1.8268 days) showed no variation corresponding to a binary system as previously reported by  \cite{bal15} (Fig.~\ref{resifig}).

\section{Spectroscopy}

Ground-based spectroscopy is used to search for possible changes in radial velocity due to binarity. Spectroscopic observations were carried out with ESPERO-a new-commissioned fiber-fed echelle spectrograph attached to the 2-m RCC telescope of Rozhen National Astronomical Observatory, Bulgaria. An Andor iKon-L CCD-camera has been used to record stellar spectra covering  the wavelength range between 4000 \AA\, and 9400 \AA\, and a typical resolving power of approximately 35\,000 at 6500 \AA. 
The mean S/N ratio was approximately 50 at 6500 \AA. Eighteen spectra have been obtained, most of them between June and September 2016 and a few more in 2017 and 2018. IRAF {\footnote{\it http://iraf.noao.edu/} standard procedures were used for bias subtracting, flat-field normalization and wavelength calibration. The final spectra were corrected to the heliocentric wavelengths.

The radial velocity have been measured by using the procedure FXCOR of IRAF. We calculated synthetic spectra of chosen spectral regions by using the code SYNSPEC (\cite{hubeny94}) and Kurucz atmosphere models (\cite{kurucz96}) in order to use them as templates for this procedure. The results are shown in Table 2. We have included two more radial velocity measurements obtained by \cite{catan14} and \cite{ewa15} in 2011 (for details see the text below).  

In order to check our system of radial velocities we have obtained spectra of RV standards. The radial velocity measured by us from one night for the standard HD\,173398 were $-25.12\pm0.76~{\rm km\,s^{-1}}$ while according \cite{soub13} the value is $-25.119\pm0.0125~{\rm km\,s^{-1}}$. For the other star, HD\,142639 we have measured $9.8\pm0.51~{\rm km\,s^{-1}}$ from three nights. \cite{fam05} gives the value $9.19\pm0.21~{\rm km\,s^{-1}}$. As it is seen our accuracy of measuring radial velocities is high and the values are very close to the published ones.

The amplitude of the radial velocity variations of about 10 km/s is much more than the radial velocity errors which supports the clear radial velocity conclusion. These results allow us to confirm that the star HD\,176843 is a possible binary star which first has been proposed by \cite{uyt11}. 
Unfortunately, our attempts to connect these variations with the period of 1.$^{d}$8268 proposed by\cite{bal15} as an orbital period did not give the reasonable result.Radial velocity measurements phased on the period of 1.8268 days for HD\,176843 are also presented in Figure 5. The careful check of the spectra for double lines was negative. We could not see any lines which are split (see Fig.4). This may lead us to conclude that the star is SB1. Furthermore, we also combined our radial velocity data with the ones obtained by \cite{catan14} and \cite{ewa15} to derive plausible orbital parameters. Cross-correlation technique was used to determine the value of rvsini by means of the adopted spectra from \cite{ewa15}.
For the analysis, the {\tt rvfit} code \citep{igl15} was applied to the radial velocity data (see Table 2) and we therefore listed the resulting parameters in Table 3. According to our calculations, these variations with the period of 34.$^{d}$14 would be suggested as a possible orbital period which did give a more reasonable result than previously suggested by \cite{bal15}.Best theoretical fit to radial velocity values on the period of 34.14 days for HD\,176843 are also plotted and presented in Figure 6.

\begin{table}
	\caption{Radial velocity measurements corresponding to the phase with standard errors}
	\begin{tabular}{lllll}
		\hline
		Date		&	HJD 245+ & Phase & RV[km/s] & RV$\rm _{err}$[km/s] \\
		\hline
		17.05.2011	& 5696.66520  & 0.071 &-26.72 & $\pm$0.56\\
		19.09.2011	& 5824.35360  & 0.811 &-17.17 & $\pm$0.48\\
		21.07.2016	& 7591.44155 & 0.570 &-14.72 & $\pm$2.54\\
		22.07.2016	& 7592.40175 & 0.598 &-14.76 & $\pm$1.77\\
		22.07.2016  & 7592.44421 & 0.599 &-14.82 & $\pm$2.35\\
		23.07.2016	& 7593.44898 & 0.629 &-14.69 & $\pm$2.14\\
		23.07.2016  & 7593.49145 & 0.630 &-14.74 & $\pm$2.27\\
		16.08.2016  & 7617.40795 & 0.331 &-16.09 & $\pm$2.43\\
		17.08.2016	& 7618.41679 & 0.360 &-15.92 & $\pm$2.19\\
		20.08.2016	& 7621.36269 & 0.446 &-16.26 & $\pm$2.64\\
		15.09.2016	& 7647.48922 & 0.212 &-19.29 & $\pm$2.39\\
		15.09.2016	& 7647.51099 & 0.212 &-19.18 & $\pm$2.71\\
		16.09.2016	& 7648.32971 & 0.236 &-19.33 & $\pm$2.42\\
		16.09.2016	& 7648.35104 & 0.237 &-19.22 & $\pm$2.13\\
		17.09.2016	& 7649.38379 & 0.267 &-19.45 & $\pm$2.10\\
		17.09.2016	& 7649.40526 & 0.268 &-19.75 & $\pm$2.48\\
		12.07.2017  & 7947.44836 & 0.998 &-26.25 & $\pm$2.85\\
		12.07.2017	& 7947.47042 & 0.998 &-26.22 & $\pm$2.81\\
		29.11.2017	& 8087.20700 & 0.091 &-25.37 & $\pm$4.08\\
		07.03.2018	& 8185.50199 & 0.970 &-24.76 & $\pm$3.95\\
		\hline
	\end{tabular}
\end{table}

\begin{figure*}
	\centering
	\includegraphics{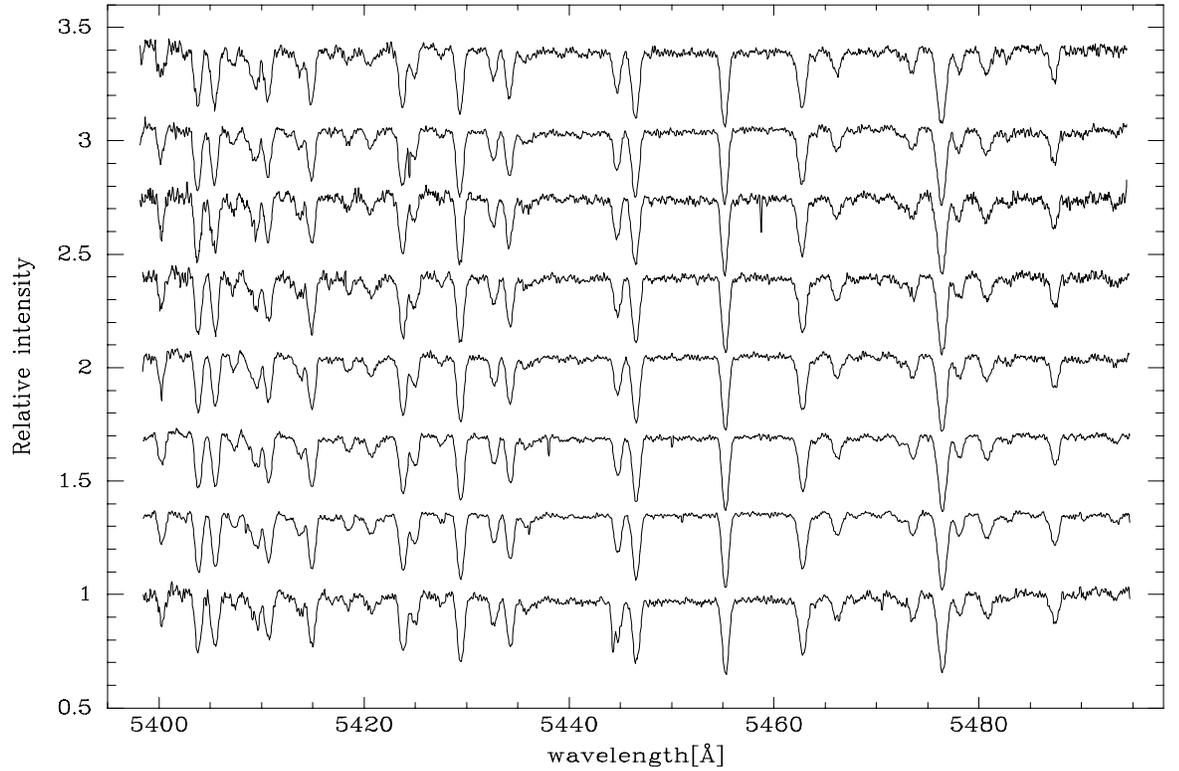}
	\caption{Normalised spectra of HD\,176843 around 5400 \AA. They are ordered by date of observation from bottom up.}
	\label{spectra}
\end{figure*}

\begin{figure*}
	\centering
	\includegraphics{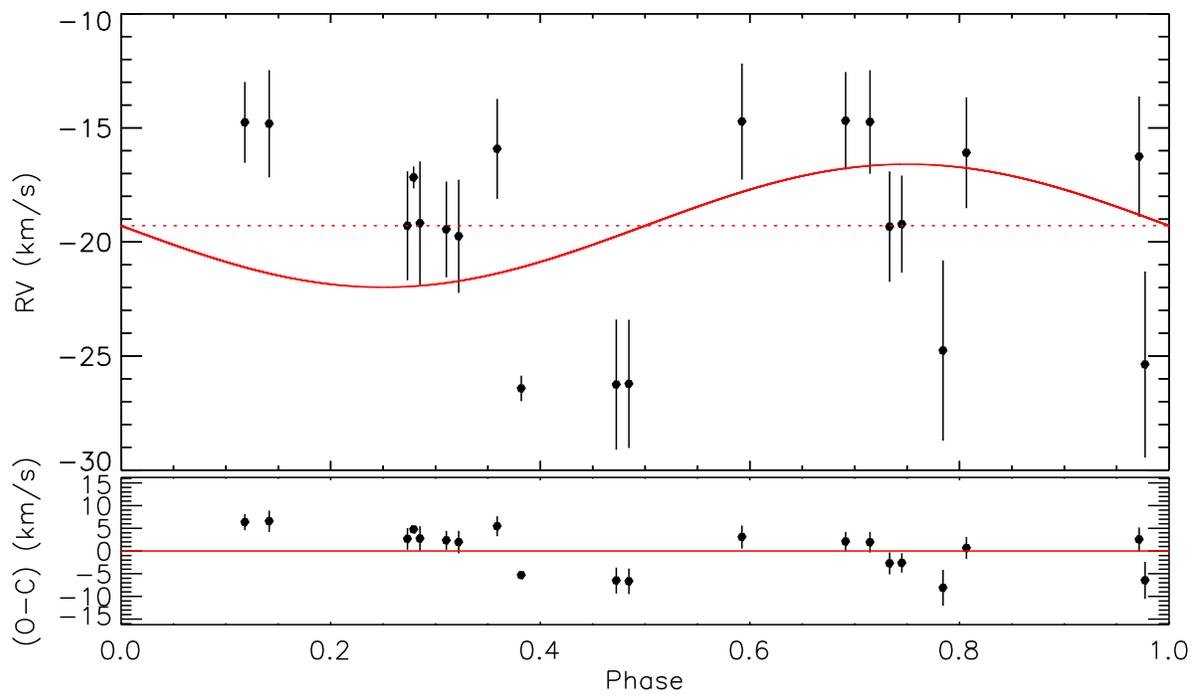}
	\caption{Radial velocity measurements phased on the period of 1.8268 days for HD\,176843.}
	\label{spectra}
\end{figure*}

\begin{figure*}
	\centering
	\includegraphics{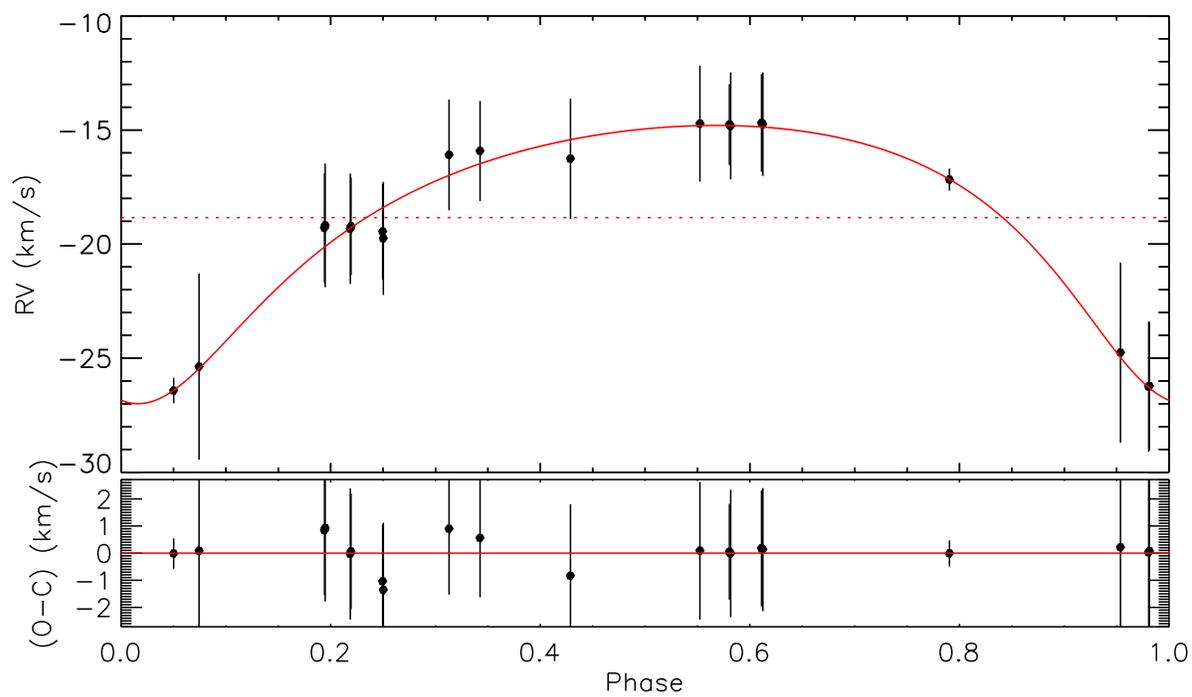}
	\caption{Best theoretical fit to radial velocity values on the period of 34.14 days for HD\,176843.}
	\label{spectra}
\end{figure*}

\begin{table}
	\caption{Orbital parameters of HD\,176843}
	\begin{tabular}{llll}
		\hline
		Parameter& Value& Error  \\
		\hline
		$P_{orb}$ (day)& 34.14 & 0.02 \\
		$T_{0}$ (HJD)  & 2455694.2532 & 0.7052 \\
		$K_{1} (km s^{-1})$ & 6.1& 0.3    \\
		$V_{\gamma} (km s^{-1})$ & -18.8&0.2   \\
		$asini$ $(10^6 km)$ & 2.66 & 0.2 \\
		$f (m)$ $(M_{\odot})$ & 0.0006 & 0.0001 \\
		e & 0.37 & 0.05   \\
		$\omega$ (deg)& 165 & 6\\
		\hline
	\end{tabular}
	
\end{table}

\section{Conclusions}

We present the results of a new study aimed to identify the variability of the marginal Am star HD\,176843 observed by the {\it Kepler} satellite. {\it Kepler} photometry and ground--based spectroscopy are used to investigate light variations of the star.
We analyzed the {\it Kepler} data to determine the possible pulsational behaviour  of HD\,176843. Frequencies were extracted from the LC data by using the software package {\tt SigSpeC}  \citep{ree07}. Our results are in a good agreement with previous analyses of the LC {\it Kepler} data reported by \cite{bal15} and we confirm that the light curve of  HD\,176843 is dominated by three modes with frequencies $f_{1}$=0.1145, $f_{2}$=0.0162 and  $f_{3}$=0.1078 d$^{-1}$.

In order to search for a possible binarity effect on the light curve of HD\,176843, we also perform the analysis of the {\it Kepler} data covering all the quarters taken in LC mode. For this purpose, all  pulsational frequencies from the data except 0.5474 d$^{-1}$ and its harmonics (1.094  d$^{-1}$ and 1.642  d$^{-1}$) were pre-whitened to test any light variation previously suggested by \cite{bal15}. Unfortunately, no significant change was derived from the residuals as shown in Figure 3.

Spectroscopic observations of HD\,176843  were obtained in 2016 and 2017 with ESPERO echelle spectrograph on the 2 m RCC telescope of the Bulgarian National Astronomical Observatory – Rozhen. Spectroscopic data analysis indicate that the star HD\,176843 clearly shows radial velocity variations and this can be interpreted as due to binarity as suggested by \cite{uyt11}. Alas, these changes in radial velocity of HD\,176843 could not be attributed to the orbital period  of 1.$^{d}$8268 proposed by \cite{bal15} (Figure 5). Also, our second attempt with additional radial velocity values and the adopted spectra \citep{ewa15} would suggest a new orbital period of 34.$^{d}$14 for these variations in  the star HD\,176843 (Figure 6).  It should be noted that no split lines were observed in the spectra (Figure 4) and it therefore seems that there is a probability of a SB1 case. Additionally, we tried to derive orbital parameters using our radial velocity measurements and two more measurements published previously by \cite{catan14} and \cite{ewa15} (corresponding to the first two RV's values in Table 2).

\section*{Acknowledgments}
The authors acknowledge the whole {\it Kepler} team for providing the unprecedented data sets that make these results possible. CU acknowledges financial support from the University of South Africa (UNISA) and the South African National Research Foundation (NRF), Grant No: 87635. This paper includes data collected by the Kepler mission. Funding for the {\it Kepler} mission is provided by the NASA Science Mission directorate. IS, II, MN acknowledge for the  partial support of the projects DN 08-1/2016 and DN 18/13-12.12.2017. MN acknowledges for the partial support of the project DFNP-103/11.05.2016.

\clearpage

\end{document}